\documentclass[journal,twocolumn,letterpaper]{IEEEJERM}

\ifCLASSINFOpdf
 
\else
 
\fi

\usepackage{times,amsmath,epsfig}
\usepackage{fancyhdr}
\usepackage{amsmath}
\usepackage{amsfonts}
\usepackage{amssymb}
\usepackage[utf8]{inputenc}
\usepackage{array}
\usepackage{bbm} 
\usepackage{graphicx}
\usepackage{url}
\usepackage{bm}
\usepackage{breqn}
\usepackage{xcolor}
\usepackage{soul}
\usepackage{amssymb}
\usepackage{cite}
\usepackage{comment}
\usepackage{url}
\usepackage{textcomp}
\usepackage{color}
\usepackage[caption=false,font=footnotesize]{subfig}

\usepackage{amsmath}
\usepackage{amssymb}
\usepackage{mathtools}
\usepackage{bm}

\usepackage{graphicx}
\usepackage{stfloats}
\usepackage{float}
\usepackage{lscape}
\usepackage{caption}

\usepackage{booktabs}
\usepackage{array}
\usepackage{multirow}
\usepackage{longtable}
\usepackage{tabulary}

\usepackage{algorithm}
\usepackage{algpseudocode}

\usepackage{pifont}

\begin{document}

\title{Rate-Aware Quantum-Inspired
Trajectory Learning for
Interference-Limited Multi-UAV
Networks}

\author{Khaoula~Khalid,
        Muhammad~Afaq, and~Zeeshan~Kaleem,~\IEEEmembership{Senior~Member,~IEEE}
}


\twocolumn[
\begin{@twocolumnfalse}

\maketitle

\begin{abstract}
Placing unmanned aerial vehicles (UAVs) as aerial base stations is non-convex and
combinatorial, so practical methods first discretize the continuous airspace into a
finite set of candidate positions---a uniform grid or clustering centroids---and optimize
over that set. Built from Euclidean distance alone, the candidate set is
\emph{channel-blind}: it ignores line-of-sight probability and the co-channel
interference that limits achievable rate under full frequency reuse. No optimizer can
select a position the candidate set omits, so good locations the geometry missed are lost
before the search begins. We propose Rate-Aware Quantum-Assisted Graph Condensation
(RA-QAGC), a two-stage framework that makes the candidate set itself the object of
optimization. The first stage constructs it by annealing a cost that embeds tier-weighted
achievable rate, mean co-channel interference, and an explicit minimum-rate penalty,
yielding a compact, mobility-feasible waypoint graph. The second stage learns
trajectories over that graph with decentralized independent $Q$-learning agents,
replacing a centralized joint-action table with one compact table per UAV. RA-QAGC
improves priority-tier throughput by 21--54\% over seven baselines, including recent
graph-neural and deep $Q$-network schemes, at no significant cost in aggregate
throughput, and the margin widens as the fleet densifies. We additionally examine the
quantum-inspired acceptance rule that recent aerial-networking work advocates, and report
that its tunnelling floor confers no measurable advantage over classical simulated
annealing at any fleet size tested, while annealing itself clearly outperforms greedy and
random placement.
\end{abstract}


\end{@twocolumnfalse}]

{
  \renewcommand{\thefootnote}{}%
  \footnotetext[1]{Khaoula~Khalid is with Department of Computer Engineering, King Fahd University of Petroleum \& Minerals (KFUPM), Dhahran 31261, Saudi Arabia (e-mail: khaoula.khaled97@gmail.com)}
  \footnotetext[2]{Muhammad~Afaq is with Department of Computer Engineering and IRC for Intelligent Secure Systems, KFUPM, Dhahran 31261, Saudi Arabia (e-mail: muhammad.afaq@kfupm.edu.sa)}
   \footnotetext[3]{Zeeshan Kaleem is with Department of Computer Engineering and Interdisciplinary Research Center for Smart Mobility and Logistics, KFUPM, Dhahran 31261, Saudi Arabia (e-mail: zeeshankaleem@gmail.com)}
}
 
\IEEEpeerreviewmaketitle

\section{Introduction}
\IEEEPARstart{T}{he} airspace below a few hundred metres is being opened to
commercial use at scale. National programs increasingly treat this low-altitude
band as an economic domain in its own right, spanning logistics, infrastructure
inspection, urban air mobility and public safety, and the communications
infrastructure needed to support it is now specified alongside the aircraft
themselves~\cite{ahmed2025lae}. Wireless connectivity is the common dependency across these applications: each assumes a link available where and when it is
needed, rather than where a tower happens to stand. That expectation has shifted
the aerial platform from a payload to be served into a network element ~\cite{11435927, apte2026survey}.
The integration of uncrewed aerial vehicles (UAVs) into next-generation wireless networks promises resilient, on-demand three-dimensional connectivity, particularly for emergency communications after terrestrial infrastructure fails. Altitude-dependent line-of-sight (LoS) advantages and rapid repositioning make aerial base stations (ABSs) attractive wherever demand is transient or infrastructure is damaged~\cite{10612836}.

Nowhere is the shift more consequential than when terrestrial infrastructure is
unavailable. After an earthquake, flood or storm, the surviving macro cells are
typically those furthest from the affected population, and restoring service by
conventional means takes days. UAV carrying a base-station payload can be
airborne within minutes and positioned directly above the demand, which is why
rapidly deployable aerial cells now appear in public-safety network
planning~\cite{apte2026survey}. The same capability addresses a less dramatic but
far more frequent case: transient demand surges at venues, construction sites and
transport hubs, where permanent capacity would stand idle for most of the
year~\cite{10612836}.
 
A single aerial cell is nonetheless both capacity-limited and a single point of
failure, so fleets rather than individual platforms are the operating assumption:
a group of aircraft can reconfigure around a loss and cover a region no one
platform could serve~\cite{alkamali2026relays}. Fleets, however, change the
problem rather than merely scaling it. Once several ABSs share spectrum under full
frequency reuse, the dominant limit on achievable rate ceases to be the distance
from a user to its serving platform and becomes the mutual interference among the
platforms themselves, which placement rules derived from coverage geometry cannot
see~\cite{zhou2025fluid}. The deployment decision therefore governs not only how
much capacity exists but who receives it, a distinction that matters acutely
when part of the population, such as first responders, medical teams or command
traffic, carries a service guarantee that the rest does not.
 
However, the 3D placement of ABS has several challenges. For instance, joint trajectory design and resource allocation is non-convex and combinatorial, and its configuration space grows exponentially with the number of ABSs and candidate positions~\cite{11733725}. Model-based approaches based on successive convex approximation and block coordinate descent converge to stationary points but require centralized iterative computation that scales poorly~\cite{11358907}. Geometric clustering reduces the search space cheaply but is \emph{channel-blind}: centroids are computed from Euclidean distances alone and ignore LoS probability, excess attenuation, and co-channel interference~\cite{10409538}. 

Deep and multi-agent reinforcement learning (DRL/MARL) dispense with an explicit channel model and adapt online, but they must still be told what the agents are permitted to do, and that action set is supplied before training rather than derived from the radio objective it is meant to serve. In practice it is a uniform grid or a fixed candidate list: the multi-agent DQN agents of~\cite{chen2026efficient} each select a base-station site from a discrete set of candidate positions, and the trajectory and association schemes of~\cite{shamsabadi2026dqn} act over a comparable pre-specified discretization of the deployment region. The consequence is a dilemma rather than a tuning choice: where the discretization is coarse the attainable policy is coarse with it, and where it is fine the joint action space grows beyond what a learner can explore within a practical episode budget. The difficulty is now explicitly acknowledged. Authors in \cite{akhtarshenas2026ppo} criticize predefined spatial grids directly and avoid them by having a centralized Proximal Policy Optimization (PPO) controller emit continuous displacements instead. That removes the discretization rather than improving it, however, and a continuous joint action space prevent the tabular decentralization on which tractability for large fleets depends. In neither case is the abstraction itself treated as a quantity to be optimized.

To address the issues in the previous works, we have following contributions:

\begin{itemize}
\item We build the deployment candidate set by minimizing a cost defined in tier-weighted achievable rate, co-channel interference, and minimum-rate violations rather than Euclidean dispersion, so the
positions retained are those that deliver rate under full-reuse interference. Condensing
the deployment space is not itself new; defining the criterion in rate and interference
terms is.

\item The resulting compact,
mobility-feasible waypoint graph is handed to decentralized independent $Q$-learning
agents, replacing the centralized joint-action table with one small table per UAV. We
further separate the effect of \emph{where} waypoints are placed from \emph{how many}
there are by comparing against uniform grids of equal and of higher cardinality.

\item We compare against seven baselines, including recent graph-neural and deep $Q$-network schemes, under a
common channel, allocation, and constraint model. We then delimit what may be credited
to the framework: ablation shows that no individual component of the condensation or
search stage produces a difference resolvable against drop-to-drop variation, and the
quantum-inspired tunnelling floor confers no measurable advantage over classical
simulated annealing at any fleet size tested.
\end{itemize}
 
\section{Related Work}
\label{sec:related}
We discuss the related works and state the shortcoming that existing schemes have not addressed. 
Because the continuous problem is intractable at scale, a large body of work compresses the environment before optimizing, almost always by clustering users and treating cluster centroids as candidate ABS positions \cite{kaleem2022learning}. For instance, classical $K$-means~\cite{1017616} partitions users by Euclidean dispersion and places one ABS at each centroid, which is the abstraction underlying the UAV-specific clustering schemes. Moreover, weighted $K$-means (W-$K$-means) modifies the centroid update so that priority users carry weight $w_{\mathrm{pr}}$ and normal users $w_{\mathrm{nr}}$, which is the cheapest way to make a geometric abstraction tier-aware and therefore the most direct test of whether tier-awareness alone, without channel awareness, suffices. But the problem is that, a centroid can sit at the geometric center of a dense cluster and still be a poor ABS position, because the users it serves are mutually interfering or shadowed. Since the downstream optimizer can only choose among the candidates the clustering produced, this blindness is not recoverable later.
In graph based condensation schemes, Zhang \emph{et al.}~\cite{zhang2023cooperative} presented Independent Q-learning with graph vision (iGCVis), who designed cooperative trajectories for multiple UAV base stations using heterogeneous graph neural networks, representing UAVs and ground users as distinct node types and their service relationships as typed edges, with a multi-agent actor--critic loop trained on the learned graph representation. The architecture is well matched to the relational structure of the problem and demonstrates that a learned graph embedding can coordinate a fleet. Its shortcoming is precisely the one at issue here: the graph encodes spatial and association relations, so the embedding is trained to predict \emph{who is near whom and who serves whom} rather than what rate a configuration achieves. Spatial proximity substitutes for instantaneous signal-to-interference and noise ratio (SINR), and in deep-fading or strongly interference-coupled regimes that substitution breaks down, which is where the reported convergence degrades. 

Reinforcement learning avoids explicit channel modeling by learning policies from interaction, and has been applied across the aerial deployment problem~\cite{kaleem2022learning, 10409538}. Shamsabadi \emph{et al.}~\cite{shamsabadi2026dqn} used a centralized deep Q-network (DQN) for joint UAV trajectory and user-association planning (JUTAP) in non-terrestrial-network-assisted systems, minimizing handover-induced disconnectivity through macro-level planning; the scheme is effective at its stated objective, but because the reward is dominated by association continuity it carries no fine-grained interference-aware mechanism and degrades in dense co-channel deployments. Cai \emph{et al.}~\cite{Cai2025MARL} applied multi-agent deep reinforcement learning to joint coverage and power-efficient user-equipment association, demonstrating that decentralized agents can trade coverage against UE power draw without a central controller; the agents, however, act over a coarse discretization of the deployment region supplied as a fixed input, so the resolution and placement of that discretization bound the attainable policy and are never themselves adapted.

The structural obstacle behind both is the joint action space. Learning over $N$ agents with per-agent action set $\mathcal{A}$ induces a joint space of cardinality $\mathcal{O}(|\mathcal{A}|^{N})$. Decentralized formulations such as independent $Q$-learning (IQL) break the exponential dependence at the cost of environmental non-stationary, and are the standard remedy. But the remedy addresses the exponent, not the base: a decentralized learner over a needlessly large or poorly sited action set still learns slowly. In all of the works above the learner is applied directly to a uniform grid or to a continuous space, so the size and siting of the action set, which is the base of that exponential, is never itself optimized.

Recently the quantum-driven approaches \cite{ranjha2026qien} has attracted the researchers attention because of it's ability to solve the hard problems \cite{11771018}. Annealing-type global search is now established as competitive with commercial solvers on combinatorial landscapes~\cite{Huang2022BenchmarkingQA,Chang2024QuantumCF}, and has been applied at aerial networking. Jeong \emph{et al.}~\cite{Jeong2025QuantumAnnealingBasedSR} cast joint user clustering, subchannel assignment, and power allocation for sum-rate maximization as a quadratic binary problem and solved it by quantum annealing, obtaining solutions competitive with iterative classical baselines at a fraction of the iteration count. The results confirmed that global search of this type is well suited to the discrete, rugged cost landscapes that aerial deployment produces. However, the quantum-inspired algorithm is applied to \emph{solving} the placement or resource-allocation problem posed over a pre-existing variable set. The candidate positions, subchannels, or discretized states are given beforehand, so the search inherits whatever abstraction was handed to it and cannot improve upon it. Similarly, the authors in~\cite{11535139} took a first step toward the abstraction itself, introducing quantum-driven state reduction for UAV trajectory optimization. That formulation, however, targeted an outage-probability objective under a single service class, screened candidate states by a criterion that did not reference achievable rate, and did not model full-reuse co-channel interference. It therefore established that state reduction is worthwhile without establishing what a reduced state space built \emph{for a rate objective under interference} should look like, which is the question taken up here.

Model-based schemes optimize a defensible objective but act on the continuous position variable and do not scale. Clustering schemes scale precisely because they compress, yet compress against a geometric surrogate that discards interference, and minimizing that surrogate more effectively does not repair it. Learning-based schemes adapt online but inherit whatever state representation they are handed and are bounded by it. Quantum and quantum-inspired schemes bring powerful global search and improved sample efficiency, but are aimed at solving the problem posed over an existing variable set, or at the replay and parameterization layers, rather than at how that variable set is constructed in the first place. Each strand, in short, treats the abstraction as given and competes on what is done with it.

Therefore, no existing framework therefore treats the reduced state space as something to be designed. What is missing is a condensation objective expressed in tier-weighted achievable rate and co-channel interference rather than Euclidean geometry, and a demonstration that a graph built under such an objective is a better substrate for decentralized learning than the
uniform grids and geometric centroids now in use. This is the gap the present work closes. We propose Rate-Aware QoS-Constrained Graph Condensation (RA-QAGC), a two-stage framework in which the abstraction is the object of optimization rather than an input to it. Stage~1
condenses the continuous deployment region into a compact, mobility-feasible waypoint graph by minimizing a communication-centric cost that embeds tier-weighted rate, mean co-channel interference, and an explicit minimum-rate penalty, searched by quantum-inspired annealing over a pre-screened candidate pool. Stage~2 learns trajectories over that graph with
decentralized IQL agents, which replaces the $\mathcal{O}(M^{N})$ joint-action table with $N$ tables of size $\mathcal{O}(M^{2})$.

\section{System Model}
\label{sec:system}
We consider the uplink of a multi-UAV-assisted wireless network spanning a continuous terrestrial region $\mathcal{A} \subset \mathbb{R}^2$. The network comprises a set $\mathcal{N} = \{1, \dots, N\}$ of $N$ Unmanned Aerial Vehicles (UAVs) acting as Aerial Base Stations (ABSs) to service a set $\mathcal{K} = \{1, \dots, K\}$ of $K$ stationary ground users (GUs). The system operates under a full-frequency reuse scheme across a total bandwidth of $B$ Hz, resulting in an interference-limited network environment.

Each ABS $n \in \mathcal{N}$ flies at a constant altitude $h_n$. At any discrete time slot $t \in \{1, \dots, T\}$, where $T$ denotes the finite mission horizon, the 3D coordinate vector of ABS $n$ is defined as $\bm{u}_n[t] = [x_n[t], y_n[t], h_n]^\top$, where $\bm{q}_n[t] = [x_n[t], y_n[t]]^\top \in \mathcal{A}$ denotes its time-varying horizontal position. The stationary GUs possess fixed coordinates $\bm{w}_k = [x_k, y_k, 0]^\top, \, \forall k \in \mathcal{K}$. To implement QoS, $\mathcal{K}$ is partitioned into two mutually exclusive subsets: priority users ($\mathcal{K}_{\text{pr}}$) requiring strict data rate guarantees, and normal users ($\mathcal{K}_{\text{nr}} = \mathcal{K} \setminus \mathcal{K}_{\text{pr}}$) processing standard best-effort traffic.
To mitigate the curse of dimensionality inherent in continuous trajectory optimization, the spatial search domain is mapped to a finite set of $M$ discrete candidate waypoints (centroids), precomputed via the RA-QAGC framework as $ \mathcal{C} = \{\bm{c}_m \in \mathbb{R}^2\}_{m=1}^{M}.$ Consequently, the horizontal positioning of each agent must satisfy $\bm{q}_n[t] \in \mathcal{C}, \, \forall n \in \mathcal{N}, \forall t$. Spatial state transitions between successive time slots $\bm{q}_n[t] \rightarrow \bm{q}_n[t+1]$ are governed by a global connectivity graph $\mathcal{G} = (\mathcal{C}, \mathcal{E})$, where a directional transition edge $(\bm{c}_m, \bm{c}_{m'}) \in \mathcal{E}$ exists if and only if: $\|\bm{c}_m - \bm{c}_{m'}\|_2 \leq v_{\max} \Delta t,$ where $v_{\max}$ is the maximum horizontal velocity of the UAV, and $\Delta t$ is the discrete slot duration.

The time-varying Euclidean distance $d_{k,n}[t]$ and the corresponding elevation angle $\theta_{k,n}[t]$ between terrestrial GU $k$ and flying ABS $n$ are expressed respectively as: $d_{k,n}[t] = \|\bm{u}_n[t] - \bm{w}_k\|_2,$ $\theta_{k,n}[t] = \arcsin\left(\frac{h_n}{d_{k,n}[t]}\right)$. Let $\theta_{k,n}^{\circ}[t] = \frac{180}{\pi}\theta_{k,n}[t]$ represent the elevation angle in degrees. The probability of establishing a Line-of-Sight (LoS) link follows a standard sigmoidal logistic distribution as $P_{\text{LoS},k,n}[t] = \frac{1}{1 + a \exp\left(-b\left[\theta_{k,n}^{\circ}[t] - a\right]\right)}$,
where $a$ and $b$ are environmental constant parameters dependent on the urban topology, and the corresponding Non-LoS (NLoS) probability is $P_{\text{NLoS},k,n}[t] = 1 - P_{\text{LoS},k,n}[t]$. The aggregate large-scale effective path loss (in dB) is represented as
\begin{equation}
\begin{split}
    L_{\text{eff},k,n}[t] = & \, K_0 + 10\alpha \log_{10}(d_{k,n}[t]) + \\ 
    & \chi_{\text{LoS}} P_{\text{LoS},k,n}[t] + \chi_{\text{NLoS}} P_{\text{NLoS},k,n}[t],
\end{split}
\end{equation}
where $K_0 = 20\log_{10}(4\pi f_c / c)$ represents the free-space path loss at a one meter reference distance for carrier frequency $f_c$. Here, $\alpha$ denotes the environment-specific path loss exponent, and $\chi_{\text{LoS}}, \chi_{\text{NLoS}}$ are state-dependent excess path loss variables assigned to LoS and NLoS conditions. Compounding large-scale attenuation with exponential small-scale Rayleigh fading variance $|f_{k,n}[t]|^2 \sim \exp(1)$ yields the total linear channel gain: $g_{k,n}[t] = 10^{-\frac{L_{\text{eff},k,n}[t]}{10}} \cdot |f_{k,n}[t]|^2$.
To maintain energy efficiency, GUs implement a fractional open-loop power control mechanism. The uplink transmission power of ground user $k$ is dynamically bounded in accordance with: $P_{\text{tx},k}[t] = \min\left(P_{\max},\; P_0 + \alpha_{\text{OL}} L_{\text{eff},k,\hat{n}_k[t]}[t] + 10\log_{10}(N_{\text{RB}})\right),$ where $P_{\max}$ is the maximum hardware transmission power threshold, $P_0$ is the target received power spectral density, $\alpha_{\text{OL}} \in [0,1]$ is the path loss compensation factor, and $N_{\text{RB}}$ denotes the number of allocated resource blocks. 

Terrestrial users associate with ABS that provides the maximum average received signal power computed as $\hat{n}_k[t] = \underset{n \in \mathcal{N}}{\arg\max}\; \left( p_{\text{tx},k}[t] \cdot g_{k,n}[t] \right),$ where $p_{\text{tx},k}[t]$ is the linear equivalent wattage of the logarithmic value $P_{\text{tx},k}[t]$. Under a full frequency reuse, the cumulative co-channel inter-cell interference power $I_{n,k}[t] = \sum_{j \in \mathcal{K}, j \neq k} p_{\text{tx},j}[t] \cdot g_{j,n}[t]$ experienced at ABS $n$, and the resulting received SINR $\gamma_k[t]$ for GU $k$ are $\gamma_k[t] = \frac{p_{\text{tx},k}[t] \cdot g_{k,\hat{n}_k[t]}[t]}{\sigma^2 + I_{\hat{n}_k[t],k}[t]}$, where $\sigma^2$ represents the additive white Gaussian noise (AWGN) thermal power.
we adopt a priority-weighted intra-cell bandwidth allocation policy in which user $k$ receives a share proportional to its tier weight $\phi_k = w_{\text{pr}}$ if $k \in \mathcal{K}_{\text{pr}}$ and $\phi_k = w_{\text{nr}}$ otherwise. The achievable uplink data rate for GU $k$ is therefore
\begin{equation}
R_k[t] = \min\!\left(\frac{\phi_k}{\sum_{j \in \mathcal{K}_{\hat{n}_k[t]}[t]} \phi_j},\; \beta_{\max}\right) B \log_2\left(1 + \gamma_k[t]\right),
\label{eq:rate}
\end{equation}
where $\beta_{\max}=0.20$ bounds the fraction of the carrier assignable to any single UE, consistent with the per-UE resource-block limits used in cellular schedulers. Setting $w_{\text{pr}}=w_{\text{nr}}$ and $\beta_{\max}=1$ recovers equal sharing. Here, where $|\mathcal{K}_n[t]| = \sum_{j \in \mathcal{K}} \mathbf{1}\{\hat{n}_j[t] = n\}$ represents the instantaneous user association cardinality of ABS $n$, evaluated using the indicator function $\mathbf{1}\{\cdot\}$.

To enforce priority, the aggregate network utility $\mathcal{T}[t]$ is formulated as a weighted sum-rate as $\mathcal{T}[t] = w_{\text{pr}} \sum_{k \in \mathcal{K}_{\text{pr}}} R_k[t] \;+\; w_{\text{nr}} \sum_{k \in \mathcal{K}_{\text{nr}}} R_k[t],$ where $w_{\text{pr}}$ and $w_{\text{nr}}$ denote the priority and normal tier weights respectively, satisfying $w_{\text{pr}} \gg w_{\text{nr}}$. The primary network objective is to maximize this utility over the global mission horizon.

\section{Problem Formulation \& Proposed RA-QAGC Framework}
\label{sec:problem}
In this section, we formulated the weighted sum-rate maximization problem by jointly optimizing the UAV trajectory and resource allocation over a finite operational mission horizon~$T$, represented as
\begin{align}
    \max_{\substack{\{\bm{q}_n[t], \\ p_{\text{tx},k}[t]\}}} & \quad \sum_{t=1}^{T} \left( w_{\text{pr}} \sum_{i \in \mathcal{K}_{\text{pr}}} R_i[t] + w_{\text{nr}} \sum_{j \in \mathcal{K}_{\text{nr}}} R_j[t] \right) \label{obj} \\
    \text{s.t.} & \quad X_{\min} \leq x_n[t] \leq X_{\max}, \quad Y_{\min} \leq y_n[t] \leq Y_{\max}, \label{c1:perimeter} \\
    & \quad \|\bm{q}_n[t+1] - \bm{q}_n[t]\|_2 \leq v_{\max} \Delta t, \quad \forall n \in \mathcal{N}, \, t < T, \label{c3:velocity} \\
    & \quad \bm{q}_n[t] \in \mathcal{C}, \quad \forall n \in \mathcal{N}, \, \forall t, \label{c4:waypoint} \\
    & \quad (\bm{q}_n[t], \bm{q}_n[t+1]) \in \mathcal{E}, \quad \forall n \in \mathcal{N}, \, t < T, \label{c5:graph} \\
    & \quad p_{\min} \leq p_{\text{tx},k}[t] \leq p_{\max}, \quad \forall k \in \mathcal{K}, \, \forall t, \label{c6:power_cap} \\
    &\quad \sum_{i \in \mathcal{K}_n[t]} B_i[t] \leq B, \quad \forall n \in \mathcal{N}, \, \forall t,\\
    & \quad R_i[t] \geq R_{i,\min}[t] \label{c9:qos}
\end{align}
The optimization problem in \eqref{obj} is supported by various constraints. Here,
Constraint \eqref{c1:perimeter} specifies that the ABS horizontal trajectories must remain strictly bounded within the designated terrestrial deployment region $\mathcal{A}$. Kinematic step bounds are enforced by \eqref{c3:velocity}. Constraints \eqref{c4:waypoint} and \eqref{c5:graph} enforce discrete waypoint restrictions and graph topology adherence during the trajectory tracking phase. Here, the ABS positions are restricted to the pre-screened condensed waypoint matrix $\mathcal{C}$, and their spatial state transitions must map across connected geometric edges defined within the valid physical edge domain $\mathcal{E}$ of the graph $\mathcal{G}_C$. The uplink transmit power is constrained by \eqref{c6:power_cap}, and \eqref{c9:qos} enforces the per-tier minimum-rate requirement that guarantees service to the priority tier. 

Optimizing multi-UAV trajectories jointly requires searching a configuration space of cardinality $\mathcal{O}(M^{N})$ in the number of candidate positions $M$ and the fleet size $N$, which is intractable for any fleet of practical size. The RA-QAGC framework decouples the problem into a sequential two-stage pipeline: (i) offline compression of the deployment space into a compact waypoint graph by annealing-based global search under a rate- and interference-aware cost, and (ii) online localized trajectory tracking by decentralized multi-agent reinforcement learning over that graph. The first stage determines \emph{where} the fleet may go; the second determines \emph{when} each ABS goes there. Our claim attaches to the cost minimized in Stage~1 and to the graph it yields, not to the mechanism used to minimize it.
 
\subsection{Stage 1: Rate-Aware Graph Condensation}
To evaluate the topological fitness of a multi-UAV spatial deployment layout, we construct a communication-centric cost function $J(\{\bm{q}_n\})$ that maps the wireless channel and co-channel interference characteristics of the layout into an energy landscape to be minimized:
\begin{equation}
    J(\{\bm{q}_n\}) = -\left(w_{\text{pr}} R_{\text{pri}} + w_{\text{nr}} R_{\text{nr}}\right) + \lambda \bar{I}_{\text{dB}} + \mu \, \Omega,
    \label{eq:qio_cost}
\end{equation}
where $R_{\text{pri}} = \sum_{i \in \mathcal{K}_{\text{pr}}} R_i$ and $R_{\text{nr}} = \sum_{j \in \mathcal{K}_{\text{nr}}} R_j$ are the aggregate rates delivered to the priority and normal tiers under the allocation rule~\eqref{eq:rate}. The second term,
\begin{equation}
\bar{I}_{\text{dB}} = \frac{1}{K} \sum_{k=1}^{K} 10 \log_{10}\!\left(I_{\hat{n}_k, k} + \epsilon\right),
\label{eq:ibar}
\end{equation}
is the mean co-channel interference across the served population expressed in decibels, with $\epsilon = 10^{-12}$ guarding the logarithm against empty-cell evaluations and $\lambda = 0.5$ setting its weight relative to the rate term. The third term,
\begin{equation}
\Omega = \frac{1}{K}\sum_{k \in \mathcal{K}}\mathbf{1}\!\left\{R_k < R_{k,\min}\right\},
\label{eq:omega}
\end{equation}
is the empirical probability of violating the per-tier minimum-rate requirement~\eqref{c9:qos}, weighted by $\mu = 400$. The term enters as a standard exterior penalty on the minimum-rate constraint~\eqref{c9:qos}, so that candidate deployments violating it are scored down in proportion to the fraction of users affected.
 
The structural distinction from geometric clustering lies in~\eqref{eq:ibar}. A $K$-means objective rewards proximity to demand without penalty, so it drives ABSs toward the densest user concentrations and, under full frequency reuse, toward one another. Because~\eqref{eq:qio_cost} charges for the interference that such a layout creates, a configuration that is attractive on coverage alone is penalized once its co-channel coupling is accounted for. The condensation criterion is in this sense defined against the same quantity the deployment is ultimately judged by.
 
\subsection{Stage 1: Search Procedure}
The cost~\eqref{eq:qio_cost} is non-convex and multimodal in $\{\bm{q}_n\}$, so it is minimized by annealing-based global search over a discrete candidate set. A configuration state $\bm{s} = [s_1, \dots, s_N]^\top$ holds coordinate indices into a pre-screened candidate set of size $N_{\text{cand}} = 200$, drawn from an initial pool of $N_0 = 1200$ uniformly sampled points. Pre-screening admits candidates in descending rate-aware score subject to a minimum separation $\delta$, so that the retained set preserves spatial diversity and spans the high-rate corridors of the whole terrain rather than collapsing onto the single densest hotspot. A successor state $\bm{s}'$ is proposed by selecting $n_{\text{swap}} = 2$ agents uniformly at random and reassigning their indices:
\begin{equation}
    s'_n = \begin{cases}
        \text{randi}(N_{\text{cand}}), & n \in \mathcal{R}_{\text{swap}}, \\
        s_n, & \text{otherwise}.
    \end{cases}
\end{equation}
Exploration is governed by an exponential cooling schedule,
\begin{equation}
    T_{\kappa+1} = \rho \, T_{\kappa}, \qquad T_0 = 1.0,
\end{equation}
with cooling index $\rho = 0.9985$ over $I_{\max} = 2500$ iterations, reaching a terminal temperature of $T_{I_{\max}} = 0.023$. Proposals are accepted under the classical Metropolis rule
\begin{equation}
P_{\text{accept}} = \min\!\left(1,\; \exp\!\left(-\Delta J / T_\kappa\right)\right), \qquad \Delta J = J(\bm{s}') - J(\bm{s}).
\label{eq:metropolis}
\end{equation}
 
A quantum-inspired variant of~\eqref{eq:metropolis}, advocated in recent aerial-networking work, floors the acceptance probability by a tunnelling term,
\begin{equation}
    P_{\text{accept}} = \max\!\left(\exp\!\left(-\frac{\Delta J}{T_\kappa}\right),\; p_{\text{tunnel}} \cdot \frac{T_\kappa}{T_0}\right),
    \label{eq:quantum_acceptance}
\end{equation}
with $p_{\text{tunnel}} = 0.05$, on the grounds that it sustains exploration across high cost barriers.  
The search returns the condensed waypoint set $\mathcal{C} = \{\bm{c}_m \in \mathbb{R}^2\}_{m=1}^{M}$, formed from the $M = 100$ highest-scoring distinct candidate positions visited in accepted configurations over the annealing trajectory, pruned to satisfy the minimum separation $\delta$. Retaining the elite set rather than only the terminating configuration matters, because the fleet must subsequently be able to \emph{move} among waypoints: a set of $N$ terminal positions admits no trajectory, whereas $\mathcal{C}$ spans the high-rate regions the annealer identified and therefore supports repositioning as demand shifts.
 
\subsection{Stage 2: Decentralized Trajectory Learning}
Restricting ABS positions to $\mathcal{C}$ converts the continuous search domain into a discrete state-action space governed by the mobility-feasible connectivity graph $\mathcal{G}_C = (\mathcal{C}, \mathcal{E})$, where $(\bm{c}_m, \bm{c}_{m'}) \in \mathcal{E}$ if and only if $\|\bm{c}_m - \bm{c}_{m'}\|_2 \leq v_{\max}\Delta t$. Trajectory execution is then modelled as a discrete-time multi-agent MDP $(\mathcal{S}, \mathcal{A}, \mathcal{P}, \mathcal{R}, \gamma)$. The joint state $\bm{s}[t] = [s_1[t], \dots, s_N[t]]^\top$ collects per-agent waypoint indices, and each agent selects a successor waypoint from its locally admissible set
\begin{equation}
a_n[t] \in \mathcal{A}_n(s_n[t]) = \left\{ m' : (\bm{c}_{s_n[t]}, \bm{c}_{m'}) \in \mathcal{E} \right\}.
\end{equation}
Assuming deterministic kinematic execution, the next state is the identity map of the joint action, $\bm{s}[t+1] = \bm{a}[t]$. The scalar reward mirrors the instantaneous tier-weighted network utility,
\begin{equation}
    r[t] = \mathcal{T}[t] = w_{\text{pr}} R_{\text{pri}}[t] + w_{\text{nr}} \left(R_{\text{total}}[t] - R_{\text{pri}}[t]\right),
\end{equation}
so that Stage~2 optimizes the same quantity Stage~1 condensed against, with the QoS penalty applied at evaluation.
 
Each agent maintains an independent tabular action-value function $Q_n(s_n, a_n)$ and updates it by temporal difference,
\begin{equation}
\begin{split}
Q_n(s_n, a_n) \leftarrow & \, Q_n(s_n, a_n) + \alpha \bigg[ r[t] + \\
& \gamma \max_{a' \in \mathcal{A}_n(s_n')} Q_n(s_n', a') - Q_n(s_n, a_n) \bigg],
\end{split}
\label{eq:q_update_split}
\end{equation}
with learning rate $\alpha = 0.25$ and discount factor $\gamma = 0.95$, selecting actions by $\epsilon$-greedy exploration. Sharing one global reward across agents aligns their incentives toward the network objective, but we note that independent learners in a shared environment face a non-stationary transition kernel, since each agent's effective dynamics change as its peers update. A shared reward mitigates this but does not remove it, and we make no convergence claim for the joint policy.
 
The benefit of the decomposition is the size of the representation. Centralized joint-action learning indexes a table of cardinality $\mathcal{O}(M^{N})$, whereas the decentralized formulation maintains $N$ tables of size $M \times |\mathcal{A}_n|$ with $|\mathcal{A}_n| \leq M$, giving $\mathcal{O}(NM^{2})$ overall and $\mathcal{O}(M^{2})$ per agent. Algorithm~\ref{alg:raqcgc} summarizes the complete pipeline.
 
\begin{algorithm}[!htbp]
\caption{RA-QAGC: Rate-Aware Quantum-Assisted Graph Condensation and Multi-UAV Trajectory Optimization}
\label{alg:raqcgc}
\begin{algorithmic}[1]
\Require UAV set $\mathcal{N}$, user set $\mathcal{K}$, initial pool $N_0=1200$, pre-screened candidates $N_{\mathrm{cand}}=200$, condensed waypoints $M=100$, cooling factor $\rho=0.9985$, initial temperature $T_0=1.0$, iterations $I_{\max}=2500$, cost weights $(\lambda,\mu)=(0.5,400)$, learning rate $\alpha=0.25$, discount factor $\gamma=0.95$
\Ensure Condensed waypoint set $\mathcal{C}$, graph $\mathcal{G}_C=(\mathcal{C},\mathcal{E})$, Q-tables $\{Q_n\}_{n\in\mathcal{N}}$
 
\Statex \textbf{Stage 1: rate-aware condensation}
\State Sample $N_0$ candidate locations; retain the top $N_{\mathrm{cand}}$ by rate-aware score subject to separation $\delta$
\State Initialize state $s$ and evaluate
\Statex \hspace{\algorithmicindent}$J(s)= -\big(w_{\mathrm{pr}}R_{\mathrm{pri}} + w_{\mathrm{nr}}R_{\mathrm{nr}}\big) + \lambda \bar{I}_{\mathrm{dB}} + \mu\,\Omega$
\State $\mathcal{P} \gets \emptyset$ \Comment{elite pool of visited positions}
\For{$\kappa \gets 0$ to $I_{\max}-1$}
    \State Select $n_{\mathrm{swap}}=2$ agents; propose $s'$ by reassigning their indices from the pre-screened set
    \State $\Delta J \gets J(s') - J(s)$
    \State $P_{\mathrm{accept}} \gets \min\!\left(1, \exp(-\Delta J / T_\kappa)\right)$
    \Statex \hspace{\algorithmicindent}\emph{(variant:} $\max(\exp(-\Delta J/T_\kappa),\, p_{\mathrm{tunnel}}T_\kappa/T_0)$\emph{)}
    \State With probability $P_{\mathrm{accept}}$: $s \gets s'$; $\mathcal{P} \gets \mathcal{P}\cup\{\bm{c}_{s_n}\}_{n\in\mathcal{N}}$
    \State $T_{\kappa+1} \gets \rho \, T_\kappa$
\EndFor
\State $\mathcal{C} \gets$ top-$M$ distinct positions in $\mathcal{P}$ by rate-aware score, subject to separation $\delta$
\State Build $\mathcal{G}_C=(\mathcal{C},\mathcal{E})$ with
\Statex \hspace{\algorithmicindent}$(c_m,c_{m'}) \in \mathcal{E} \iff \|c_m-c_{m'}\|_2 \le v_{\max}\Delta t$
 
\Statex \textbf{Stage 2: decentralized trajectory learning}
\For{each UAV $n \in \mathcal{N}$}
    \State Initialize $Q_n(s_n,a_n)$ over waypoint indices and admissible neighbours
\EndFor
\For{$t \gets 1$ to $T$}
    \For{each UAV $n \in \mathcal{N}$}
        \State Select $a_n[t]$ by $\epsilon$-greedy exploration over $\mathcal{A}_n(s_n[t])$
    \EndFor
    \State Execute actions; observe
    \Statex \hspace{\algorithmicindent}$r[t]= w_{\mathrm{pr}}R_{\mathrm{pri}}[t] + w_{\mathrm{nr}}\big(R_{\mathrm{total}}[t]-R_{\mathrm{pri}}[t]\big)$
    \For{each UAV $n \in \mathcal{N}$}
        \State $Q_n(s_n,a_n) \gets Q_n(s_n,a_n) + \alpha\Big[r[t] + \gamma \max_{a' \in \mathcal{A}_n(s_n')} Q_n(s_n',a') - Q_n(s_n,a_n)\Big]$
    \EndFor
\EndFor
\State \Return $\mathcal{C}$, $\mathcal{G}_C$, $\{Q_n\}_{n\in\mathcal{N}}$
\end{algorithmic}
\end{algorithm}

\section{Simulation Results and Discussion}
\label{sec:results}
 
The network occupies a $1000\times1000$~m$^2$ region. Unless stated otherwise the main operating point is $N=6$ ABSs and $K=150$ ground users, with demand organized into $8$ clusters of unequal population and the priority tier concentrated in a subset of those clusters rather than spread uniformly. This regime is deliberate: when the number of demand clusters exceeds the fleet size, the deployment must decide \emph{which} demand to serve, and it is precisely there that a tier- and channel-aware objective can differ from a geometric one. Parameters are listed in Table~\ref{tab:sim_parameters}. All results are averaged over independent user drops with error bars denoting one standard deviation.
 
All schemes are scored under an identical channel model, association rule, scheduler, feasibility constraint set, and evaluation budget, so that the reported differences reflect ABS placement alone.
 
\begin{table}[!t]
\centering
\caption{Simulation Parameters}
\label{tab:sim_parameters}
\scriptsize
\setlength{\tabcolsep}{3pt}
\begin{tabular}{@{}ll@{}}
\toprule
\textbf{Parameter} & \textbf{Value} \\
\midrule
Terrain / ABS altitude & $1000\times1000$~m$^2$ / $100$~m \\
ABSs $N$ / users $K$ & $6$ (swept $3$--$12$) / $150$ (swept $100$--$200$) \\
Demand clusters / priority ratio & $8$ / $30\%$, concentrated \\
Min.\ inter-ABS separation $d_{\min}$ & $120$~m (hard constraint) \\
Carrier $f_c$ / bandwidth $B$ & $2.0$~GHz / $20$~MHz \\
LoS environment $(a,b)$ & Urban $(9.61,\,0.16)$ \\
Path-loss exponent $\alpha$ & $2.5$ \\
Excess attenuation & $\chi_{\text{LoS}}=0$, $\chi_{\text{NLoS}}=20$~dB \\
Small-scale fading & Rayleigh, $|f_{k,n}|^2\sim\exp(1)$ \\
Noise floor $\sigma^2$ & $-90$~dBm \\
Open-loop $P_0$ / $\alpha_{\text{OL}}$ / $N_{\text{RB}}$ & $-90$~dBm / $0.8$ / $100$ \\
UE power range & $[-10,\,20]$~dBm \\
Tier weights $(w_{\text{pr}},w_{\text{nr}})$ & $(2.0,\,1.0)$ \\
Per-UE carrier cap $\beta_{\max}$ & $0.20$ \\
Min.\ rates $(R^{\text{pr}}_{\min},R^{\text{nr}}_{\min})$ & $(0.30,\,0.10)$~Mbps \\
Backhaul cap per ABS & $40$~Mbps \\
Cost weights $(\lambda,\mu)$ & $(0.5,\,400)$ \\
Candidates $N_0$ / $N_{\text{cand}}$ / $M$ & $1200$ / $200$ / $100$ \\
Annealing $(T_0,\rho,I_{\max})$ & $(1.0,\,0.9985,\,2500)$ \\
Tunnelling constant $p_{\text{tun}}$ (variant only) & $0.05$ \\
IQL $(\alpha,\gamma)$ & $(0.25,\,0.95)$ \\
\bottomrule
\end{tabular}
\end{table}
 
The comparison spans three groups. \emph{Geometric abstractions:} random feasible placement subject only to the separation constraint, serving as the uninformed control; classical $K$-means~\cite{1017616}, which places one ABS at each Euclidean centroid; and weighted $K$-means (W-$K$-means), which weights the centroid update by tier and therefore tests whether tier-awareness alone, without channel-awareness, suffices. \emph{Condensation variants:} graphically condensed (GC)-$K$-means and GC-(singal-to-noise ratio (SNR)) based positioning (GC-SNRP) retain the Stage-1/Stage-2 pipeline but substitute the condensation criterion, scoring candidates by geometric coverage and by isolated received SNR respectively, which isolates the contribution of the criterion from that of the pipeline. \emph{Learning-based schemes:} iGCVis follows the heterogeneous graph-neural trajectory design of~\cite{zhang2023cooperative}, and DQN-JUTAP follows the centralized deep $Q$-network joint trajectory and association planner of~\cite{shamsabadi2026dqn}.
 
\begin{figure*}[!t]
\centering
\includegraphics[width=\textwidth]{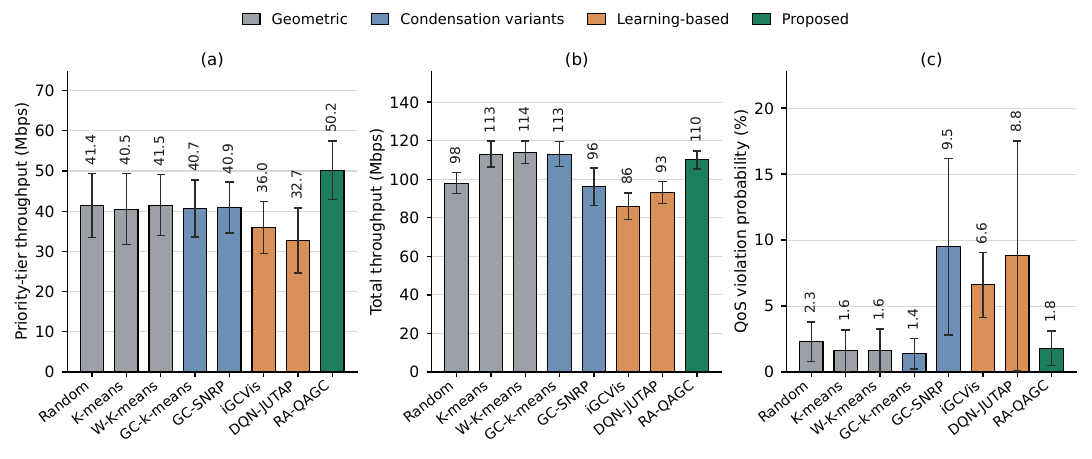}
\caption{Main comparison over 20 independent deployments ($N=6$, $K=150$): (a) priority-tier throughput, (b) aggregate throughput, (c) QoS violation probability. Error bars denote one standard deviation.}
\label{fig:main}
\end{figure*}

Fig.~\ref{fig:main} compares RA-QAGC against the seven baselines over 20 independent deployments, and Table~\ref{tab:main} reports the corresponding paired Wilcoxon signed-rank tests, where $\Delta$ is the mean \emph{relative} difference in RA-QAGC's favour; $p$ is a
two-sided paired Wilcoxon signed-rank test on the per-drop differences. The
non-significant aggregate entries ($p>0.1$) indicate no detectable throughput cost.
 
\begin{table}[!t]
\centering
\caption{RA-QAGC versus baselines: mean difference and paired Wilcoxon $p$-value over 20 deployments.}
\label{tab:main}
\footnotesize
\setlength{\tabcolsep}{3.5pt}
\begin{tabular}{@{}lcccc@{}}
\toprule
\multirow{2}{*}{\textbf{Baseline}} & \multicolumn{2}{c}{\textbf{Priority throughput}} & \multicolumn{2}{c}{\textbf{Aggregate throughput}} \\
\cmidrule(lr){2-3}\cmidrule(lr){4-5}
 & $\Delta$ (\%) & $p$ & $\Delta$ (\%) & $p$ \\
\midrule
Random          & $+21.4$ & $1.7\!\times\!10^{-4}$ & $+12.3$ & $1.9\!\times\!10^{-6}$ \\
$K$-means       & $+23.9$ & $1.7\!\times\!10^{-4}$ & $-2.6$  & $0.14$ \\
Weighted $K$-means & $+20.9$ & $1.3\!\times\!10^{-4}$ & $-3.5$ & $0.11$ \\
GC-$K$-means    & $+23.3$ & $3.2\!\times\!10^{-4}$ & $-3.1$  & $0.11$ \\
GC-SNRP         & $+22.7$ & $1.7\!\times\!10^{-4}$ & $+14.6$ & $8.2\!\times\!10^{-5}$ \\
iGCVis~\cite{zhang2023cooperative} & $+39.5$ & $4.8\!\times\!10^{-5}$ & $+27.6$ & $3.8\!\times\!10^{-6}$ \\
DQN-JUTAP~\cite{shamsabadi2026dqn} & $+53.8$ & $5.7\!\times\!10^{-6}$ & $+18.5$ & $3.8\!\times\!10^{-6}$ \\
\bottomrule
\end{tabular}
\end{table}

\begin{figure}[!htbp]
\centering
\subfloat[]{%
  \includegraphics[width=0.3\textwidth]{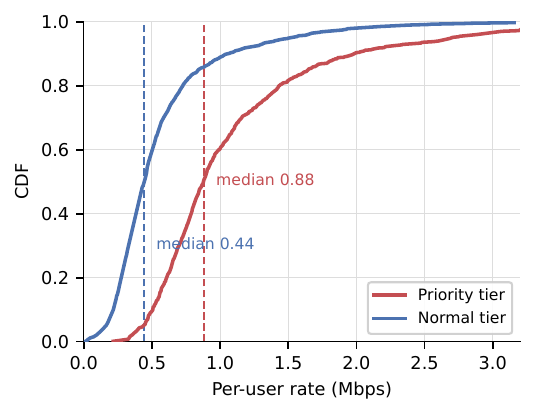}%
  \label{fig:cdf_tier}}
\hfil
\subfloat[]{%
  \includegraphics[width=0.3\textwidth]{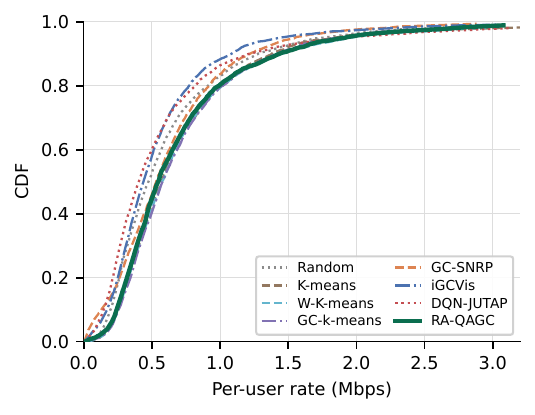}%
  \label{fig:cdf_schemes}}
\caption{Per-user rate CDFs: (a) by service tier under RA-QAGC, (b) across all evaluated schemes.}
\label{fig:cdf}
\end{figure}

\begin{figure}[!t]
\centering
\includegraphics[width=0.9\linewidth]{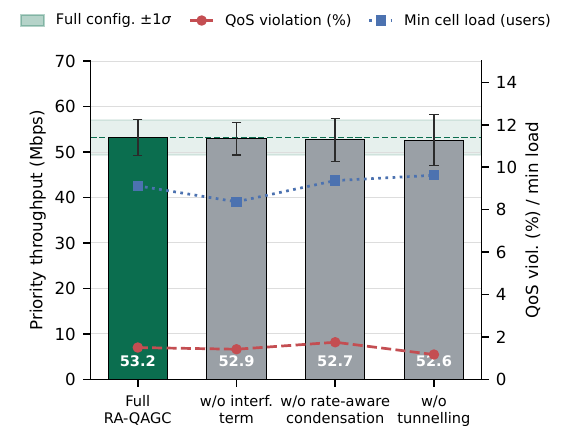}
\caption{Ablation of RA-QAGC components. Bars denote priority throughput (left axis)
with one standard deviation; the shaded band is the full configuration $\pm1\sigma$.
Lines denote QoS violation probability and minimum cell load (right axis). Every
ablated variant falls within the band, so no resolvable benefit is claimed for any
individual component at this operating point.}
\label{fig:ablation}
\end{figure}

\begin{figure}[!t]
\centering
\subfloat[]{%
  \includegraphics[width=0.40\textwidth]{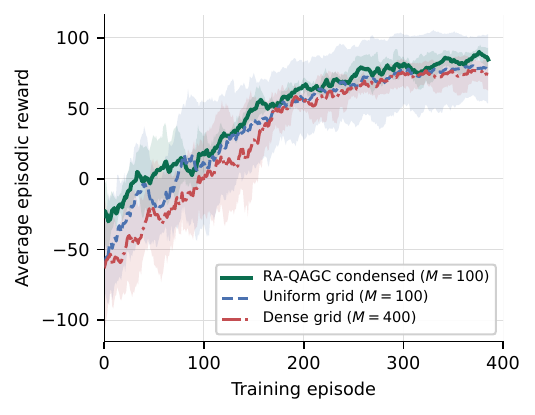}%
  \label{fig:stage2_learn}}
\hfil
\subfloat[]{%
  \includegraphics[width=0.40\textwidth]{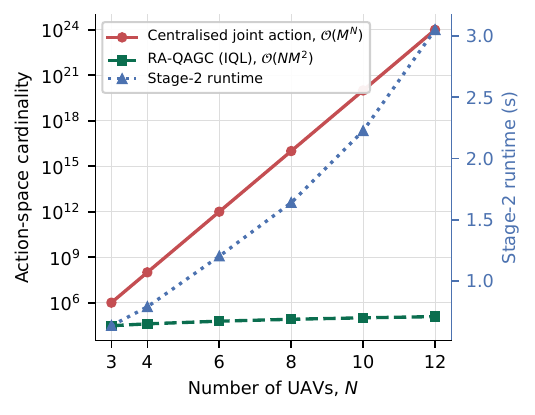}%
  \label{fig:stage2_cplx}}
\caption{Stage-2 behavior: (a) IQL learning curves on the condensed graph versus uniform grids of equal and higher cardinality; (b) action-space cardinality of centralized joint-action learning versus decentralized IQL, with measured runtime overlaid.}
\label{fig:stage2}
\end{figure}

RA-QAGC delivers $50.2$~Mbps of priority-tier throughput against $32.7$--$41.5$~Mbps for the baselines, a gain of $21$--$54\%$ that is significant at $p<5\times10^{-4}$ against every comparator. Against the three strongest geometric methods the aggregate throughput difference is $-2.6\%$ to $-3.5\%$ and \emph{not} statistically significant ($p=0.11$--$0.14$), while against the remaining four it is significantly positive. The design therefore reallocates capacity toward the priority tier without paying for it in total capacity.

Aggregate throughput clusters tightly across all methods that avoid pathological layouts, so the axis on which deployment strategies actually separate is tier-differentiated service rather than total capacity. The three schemes that fall visibly below that cluster do so for reasons traceable to their abstractions. GC-SNRP ranks candidates by isolated received SNR and is therefore blind to the coupling that full frequency reuse creates, selecting layouts whose members score well individually and poorly together; it exhibits the highest QoS violation probability at $9.5\%$. DQN-JUTAP optimizes a reward dominated by association continuity, which is the correct objective for handover minimization but leaves it without a fine-grained interference-aware mechanism. iGCVis encodes spatial and association relations rather than an explicit rate objective, so spatial proximity substitutes for instantaneous SNR.
 
Fig.~\ref{fig:cdf}(a) shows the per-user rate distributions by tier. The priority-tier median is roughly $2.3\times$ the normal-tier median, while the normal-tier distribution remains strictly positive at the 5th percentile, confirming that differentiation is achieved without starving best-effort users. Fig.~\ref{fig:cdf}(b) shows the aggregate distributions across schemes, which separate into the same two groups: the geometric abstractions and RA-QAGC are nearly coincident, which is the expected outcome rather than a null result, since under a capped weighted-sharing rule the per-cell sum rate is approximately $B$ times the cell's mean spectral efficiency and placement moves that quantity only weakly. The aggregate view therefore identifies which abstractions fail but cannot distinguish among those that do not; that separation appears only in the tier-resolved view of Fig.~\ref{fig:cdf}(a).

 
The ablation is performed to verify the impact of different blocks. Fig.~\ref{fig:ablation} removes one component at a time and re-scores the resulting deployment on the full metric set. The full configuration attains $53.2$~Mbps of priority throughput at a minimum cell occupancy of $9.1$ users. Replacing the rate-aware condensation criterion with channel-blind geometric scoring gives $52.7$~Mbps; removing the interference term from~\eqref{eq:qio_cost} gives $52.9$~Mbps; replacing the Metropolis rule by the tunnelling variant~\eqref{eq:quantum_acceptance} gives $52.6$~Mbps. All three differences fall within one standard deviation across drops, and we claim no resolvable benefit for any of them at this operating point. The interpretation we favor is that once the candidate pool spans the terrain, the \emph{objective} being optimized determines the attained deployment far more than the criterion used to pre-screen candidates or the mechanism used to search among them. The value of rate-aware condensation accordingly lies in Stage-2 tractability and sample efficiency rather than in the quality of the Stage-1 optimum.

Fig.~\ref{fig:stage2}(a) shows IQL learning curves on three action spaces. Learning on the RA-QAGC condensed graph converges to an average episodic reward of $84.5$, against $78.8$ on a uniform grid of equal cardinality and $74.4$ on a dense grid of $M=400$ waypoints, and it also begins from a markedly better initial policy. The comparison against the dense grid quantifies the curse of dimensionality directly: quadrupling the action space lowers attained reward by $12\%$ within the same episode budget. The comparison against the equal-cardinality uniform grid is the more informative of the two, since holding the action-set size fixed isolates the benefit attributable to \emph{where} waypoints are placed from the benefit of simply having fewer of them; it yields a smaller but consistent $7\%$ advantage.

Fig.~\ref{fig:stage2}(b) contrasts complexity. Centralized joint-action learning over $N$ agents on $M$ waypoints indexes a table of cardinality $\mathcal{O}(M^{N})$, exceeding $10^{24}$ entries at $M=100$, $N=12$. The decentralized formulation maintains $N$ tables of size $M\times|\mathcal{A}_n|$ with $|\mathcal{A}_n|\leq M$, giving $\mathcal{O}(NM^{2})$, or $1.2\times10^5$ entries under the same configuration. Measured Stage-2 runtime grows from $0.64$~s at $N=3$ to $3.05$~s at $N=12$, approximately linear in $N$ and consistent with the analysis. The price is the loss of joint-action coordination guarantees, which the shared global reward mitigates but does not eliminate.
 
\begin{figure*}[!t]
\centering
\subfloat[]{%
  \includegraphics[width=0.4\textwidth]{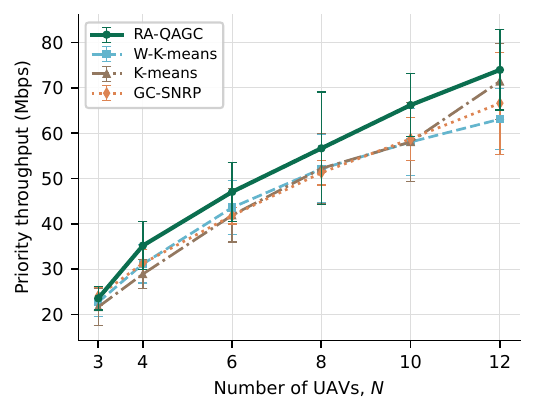}%
  \label{fig:scal_pri}}
\hfil
\subfloat[]{%
  \includegraphics[width=0.4\textwidth]{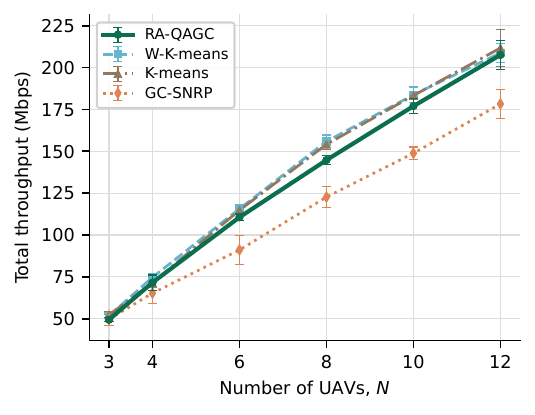}%
  \label{fig:scal_tot}}
\caption{Scalability in fleet size at $K=200$: (a) priority-tier throughput, (b) aggregate throughput.}
\label{fig:scal}
\end{figure*}

Fig.~\ref{fig:scal} sweeps fleet size from $N=3$ to $N=12$ at $K=200$ with $10$ demand clusters. The priority-tier advantage of RA-QAGC persists at every fleet size and widens as the fleet grows, from $+4\%$ over weighted $K$-means at $N=3$ to $+17\%$ at $N=12$. Aggregate throughput remains statistically indistinguishable from the geometric baselines throughout. The widening margin is consistent with the interference-limited interpretation: as the fleet densifies, co-channel coupling becomes the binding constraint, and the tier-weighted, QoS-penalized objective allocates the scarce interference-free capacity to the users the operator has designated as critical.

\section{Conclusion}
\label{sec:conclusion}
This work treated the state abstraction that precedes multi-UAV deployment optimization
as a design variable rather than a fixed geometric input. The proposed framework
condenses the deployment region into a compact, mobility-feasible waypoint graph under a
cost combining tier-weighted rate, co-channel interference, and a minimum-rate penalty,
then learns trajectories over it with decentralized independent Q-learning agents,
replacing one centralized joint-action table with a small table per aerial base station. Three findings stand out. Where waypoints sit matters more than how many there are: learning on the condensed graph gains seven percent in episodic reward over a uniform
grid of the same size, and twelve percent over one four times denser. The framework
shifts capacity toward the priority tier without paying for it in total throughput,
improving priority-tier throughput by twenty-one to fifty-four percent over seven
baselines at no detectable aggregate cost, with the margin widening as the fleet
densifies. 
\ifCLASSOPTIONcaptionsoff
\newpage
\fi
\bibliographystyle{IEEEtran}
\bibliography{Biblio}

\begin{IEEEbiographynophoto}{Khaoula Khalid}~received Engineering Degree in Computer Programming and Specific Applications from the National School of Applied Sciences of Marrakech, Morocco, in 2021. She earned  Master of Science (M.S.) degree in Computer Engineering from King Fahd University of Petroleum and Minerals (KFUPM), Dhahran, Saudi Arabia, in 2026. Currently, she is pursuing her Ph.D. degree in Computer Science, specializing in Artificial Intelligence, at the National School of Applied Sciences of Kenitra (ENSA-K), Morocco.
Her research interests include wireless communications, Unmanned Aerial Vehicles (UAVs), and advanced computational intelligence. Specifically, her work focuses on vehicle path planning, trajectory optimization, and intelligent frameworks combining reinforcement learning, machine learning, and quantum computing.
\end{IEEEbiographynophoto}

\begin{IEEEbiographynophoto}{Muhammad Afaq}~ received a B.S. degree in Electrical Engineering from the University of Eng. and Technology, Pakistan in 2007. He received an MS degree in Electrical Engineering with an emphasis on Telecom from Blekinge Institute of Technology
(Sweden) in 2010 and a Ph.D. degree in Computer Engineering from Jeju National University (Korea) in 2017. Currently, he is working as an Associate Professor at the Department of Computer Engineering, King Fahd University of Petroleum and Minerals, Saudi Arabia. His research interests are cloud
computing, SDN, NFV, computer networks and protocols, and machine learning. 
\end{IEEEbiographynophoto}

\begin{IEEEbiographynophoto}{Zeeshan Kaleem}~(Senior Member) is serving as an Associate Professor in the Computer Engineering Department, King Fahd University of Petroleum and Minerals (KFUPM), Saudi Arabia. Prior to joining KFUPM he served for 8 Years at COMSATS University Islamabad. He received his BS in Electrical Engineering from University of Engineering and Technology, Peshawar in 2007. He received MS and Ph.D. in Electronics Engineering from Hanyang University, and Inha University, South Korea in 2010 and 2016, respectively. Dr. Zeeshan consecutively received the National Research Productivity Award (RPA) awards from the Pakistan Council of Science and Technology (PSCT) in 2017 and 2018. We won the Runner-up Award in the National Hackathon 23 competition for Project to develop Drone Detection system. He won the Higher Education Commission (HEC) Best Innovator Award in 2017, with a single award from all over Pakistan. He received the 2021 Top Reviewer Recognition Award for IEEE Transactions on Vehicular Technology. He has published 100+ technical journal papers, including 21 as 1st author papers, books, book chapters, and conference papers in reputable journals/venues, and holds 21 US and Korean patents. He has also received research grants of around 90k US\$. He is a co-recipient of the best research proposal award from SK Telecom, Korea. He is currently serving as Technical Editor of several prestigious Journals/Magazines like \textit{IEEE Transactions on Vehicular Technology}, \textit{IEEE Transactions on Network and Service Management}, \textit{Elsevier Computer and Electrical Engineering}, \textit{Springer Nature Wireless Personal Communications}, \textit{Human-centric Computing and Information Sciences}, \textit{Journal of Information Processing Systems}, and \textit{Frontiers in Communications and Networks}. He has served/serving as Guest Editor for special issues in \textit{IEEE Wireless Communications}, \textit{IEEE Communications Magazine}, \textit{IEEE Access}, \textit{Sensors}, \textit{IEEE/KICS Journal of Communications and Networks}, and \textit{Physical Communications}, and served as a Track Chair in VTC-Fall 2024 and VTC-Spring 2025. He also regularly serves as TPC Member for world-distinguished conferences like IEEE Globecom, IEEE VTC, IEEE ICC, and IEEE PIMRC.
\end{IEEEbiographynophoto}
\end{document}